\documentclass[11pt]{article}
\usepackage{jcappub}

\usepackage{graphicx}
\usepackage{dcolumn}
\usepackage{bm}
\usepackage{graphics} 
\usepackage{epsfig} 
\usepackage{mathptmx} 
\usepackage{float}
\usepackage{flushend}
\usepackage{xcolor}
\usepackage{subcaption}
\usepackage{mwe}

\begin{document}

\title{Gravitational Lensing of the Cosmic Neutrino Background}

\author[a]{Joshua Yao-Yu Lin}
\affiliation[a]{Department of Physics, University of Illinois at Urbana-Champaign, 1110 West Green Street, Urbana, IL 61801, USA}
\author[a,b]{Gilbert Holder}
\affiliation[b]{Department of Astronomy, University of Illinois at Urbana-Champaign, 1002 West Green Street, Urbana, IL 61801, USA}
\emailAdd{yaoyuyl2@illinois.edu}
\emailAdd{gholder@illinois.edu}

\date{\today}
\abstract{
We study gravitational lensing of the cosmic neutrino background. This signal is undetectable for the foreseeable future, but there is a rich trove of information available. At least some of the neutrinos from the early universe will be non-relativistic today, with a closer surface of last scattering (compared to the cosmic microwave background) and with larger angles of deflection. Lensing of massive neutrinos is strongly chromatic: both the amplitude of lensing and the cosmic time at which the potential is traversed depend on neutrino momentum, in principle giving access to our entire causal volume, not restricted to the light cone. As a concrete example, we focus on the case where the cosmic neutrino background would be strongly lensed when passing through halos of galaxy clusters and galaxies. We calculate the Einstein radius for cosmic neutrinos and investigate the impact of neutrino mass. 
}

\maketitle


\section{Introduction}

In the usual thermal history of the early universe, cosmic neutrinos, also known as relic neutrinos, decouple just seconds after the Big Bang \cite{2005NuPhB.729..221M}. These neutrinos tremendously outnumber baryons in the universe; however, due to their very weak interactions with ordinary matter, no direct detection has yet been achieved \cite{2017IJMPE..2640008F}. Several papers have discussed the possibility of detecting cosmic neutrinos and their anisotropy \cite{PhysRevD.90.073006,2007JCAP...01..014M}. This will be tremendously challenging and is unlikely to be achieved soon, but there is at present no fundamental impediment to such detection. 

Since there is anisotropy in the cosmic neutrino background (C${\nu}$B) \cite{hannestad2010cosmic}, it should ultimately be possible to detect gravitational lensing of the cosmic neutrino background. As we will show below, the non-zero mass of at least two neutrino types leads to several effects that make the cosmic neutrino background and its gravitational lensing a unique source of information on the growth of structure in the universe that can be used to probe the three-dimensional structure of the universe and its time evolution, rather than the usual restriction of being on the past light-cone.

Gravitational lensing of the C${\nu}$B is similar to cosmic microwave background (CMB) lensing \cite{2006PhR...429....1L} in many respects. We show below that there are several important differences arising from the fact that neutrinos are massive \cite{2006PhR...429..307L,petcov2013nature}. Although the individual masses of neutrinos are so far unknown, we have constraints from several measurements \cite{2017PhRvD..95i6014C} that suggest that neutrino masses are expected to be in the range of several tens of meV, making the rest mass energy of many of these neutrinos larger than their thermal energy. 

Since neutrinos are massive, they must travel more slowly than light. The last scattering surface (LSS) for massive neutrinos is, therefore, closer to us than the LSS for CMB photons, by an amount that depends on the momentum of the neutrinos being measured and their mass \cite{2009PhRvL.103q1301D}. 
The background neutrinos  passing through our detectors today are physically sourced from a different three-dimensional location in space than the background photons. In the absence of gravitational deflections, those locations simply correspond to different physical locations along a single line of sight,  depending on the mass and momentum of each neutrino. Furthermore, they
pass through the intervening structures in the universe at a time that depends on neutrino mass and momentum, allowing neutrinos of different momenta to track the time evolution of individual structures. Finally, the amount of deflection also depends on the neutrino mass and momentum, allowing the same lens to be probed at a variety of impact parameters, all using neutrinos from the big bang.

In total, this presents a remarkable cache of information: the three-dimensional gravitational potential of the observable universe as a movie, limited only by causality and the poor prospects for imminent detection.

 In this work, we will focus on the simple case of strong lensing \cite{schneider1992gravitational}, as it captures many of the important physical effects, and we use a thin lens approximation, as is often done for photons.  For all the calculations in this paper, we assume a flat $\Lambda \textrm{CDM}$ expansion history and use Planck15 \cite{2016AA...594A..13P} for all cosmological parameters ($H_0 = 67.7$ km/s/Mpc , $\Omega_m = 0.307$).   
The neutrino background temperature today is $T_{\nu} = 1.95K = 1.68 \times 10^{-4}$ eV.

\section{Lensing properties of massive cosmic neutrinos}

Massless neutrinos would follow the same lensing equations as photons, as would sufficiently energetic neutrinos. For photons passing a point mass gravitational lens, the Einstein radius is given by $\theta_{E, \gamma}^{\rm PT} = \sqrt[]{\frac{4 G M}{c^2} \frac{D_{LS}}{D_S D_L}}$, where $D_S$ is the comoving distance from observer to the source (surface of last scattering), $D_L$ is the comoving distance from observer to the lens, and $D_{LS}$ is the distance between lens and source. For a singular isothermal sphere (SIS) lens, the Einstein radius is $\theta_{E, \gamma}^{\rm SIS} = \frac{4 \pi \sigma_v^2}{c^2} \frac{D_{LS}}{D_S}$, where $\sigma_v$ represents the velocity dispersion of the lens galaxy.  

For massive neutrinos, there are several differences from the massless case. First, the massive neutrinos no longer propagate at the speed of light $c$. Instead, they propagate with a time-varying velocity. We define the  neutrino momentum today as $p_0$; the momentum of neutrinos at earlier times, when the scale factor of the universe relative to today is $a(t)$, follows $a \times p(a) = p_0$. 

The momentum distribution of cosmic neutrinos follows a Fermi-Dirac distribution that is established in the first seconds after the big bang, when cosmic neutrinos were still ultra-relativistic. We can therefore follow previous work \cite{2009PhRvL.103q1301D}, assuming zero chemical potential and no additional heating from electron-positron annihilation \cite{1992PhRvD..46.3372D}, to determine the momentum probability distribution:

\begin{equation} \label{eq:FermiDirac}
\frac{dP}{dp_0} = \frac{2}{3 \zeta(3) k_B^3 T_{\nu}^3  } \frac{p_0^2 / c^3}{\mathit{e}^{p_0/k_BT_{\nu}} + 1}
\end{equation}

where $k_B$ is the Boltzmann constant. The  neutrino speed today is related to its momentum today as $p_0 = \gamma_0 m_{\nu} v_0$. The reshifted velocity can then be expressed as a function of scale factor $a$ and $v_0$. 

\begin{equation}
v(a) = \frac{v_0}{\sqrt[]{a^2 + \frac{v_0^2}{c^2} (1-a^2)}}
\end{equation}

For a neutrino of 
a particular mass and momentum today, 
we define the velocity of the neutrino
as it passes through a lens when the universe has a particular scale factor as $v_{lens}(a)$. For a neutrino with that speed, the angle of deflection as it passes a mass distribution would be 
 \cite{2014ApJ...780..158P}

\begin{equation}
\alpha (R) = \frac{4GM(R)}{R c^2} \frac{c^2 + v_{lens}^2}{2 v_{lens}^2}   \ ,
\end{equation}
where $\alpha (R)$ represents the angle of deflection of a neutrino that passes a distance $R$ from the center of mass, and $M(R)$ represents the mass interior to that radius. While very large angles of bending angle would require ray tracing simulations, we focus here on deflections that are still degree-scale rather than radian-scale. This is the Born approximation, and it definitely breaks down when the incoming neutrino is nearly bound. The deflections here are large compared to the case of photons, but for typical cosmic neutrinos they are still a small fraction of a radian. Hence, the thin lens approximation could give us first order estimation about the Einstein Radius. Using the lens equation \cite{1994A&A...284..285K} , we can get the Einstein radius for a point mass:

\begin{equation}
\theta_E^{\rm PT} = \bigg[  \frac{4 G M }{c^2} \Big(\frac{c^2 + v_{lens}^2}{2 v_{lens}^2} \Big) \frac{D_{LS}}{D_S  D_L} \bigg]^\frac{1}{2} \ ,
\end{equation}
and for a singular isothermal sphere (SIS):
\begin{equation}
\theta_E^{\rm SIS} = \frac{4 \pi \sigma_v^2}{c^2} \Big(\frac{c^2 + v_{lens}^2}{2 v_{lens}^2} \Big) \frac{D_{LS}(v_0, D_L)}{D_S(v_0)} \ ,
\end{equation}
where

\begin{equation} \label{a_lens_D_L_relation}
D_L = \int^{1}_{a_{lens}} \frac{da}{a^2 H(a)} v(a)  \ ,
\end{equation} 

\begin{equation}
D_S (v_0) = \int^{1}_{a_{s}} \frac{da}{a^2 H(a)} v(a) \ ,
\end{equation}
and 
\begin{equation}
D_{LS}(v_0, D_L) = D_S(v_0) - D_L \ .
\end{equation}

\begin{figure}[H]
	\centering
    \includegraphics[scale=0.4
    ]{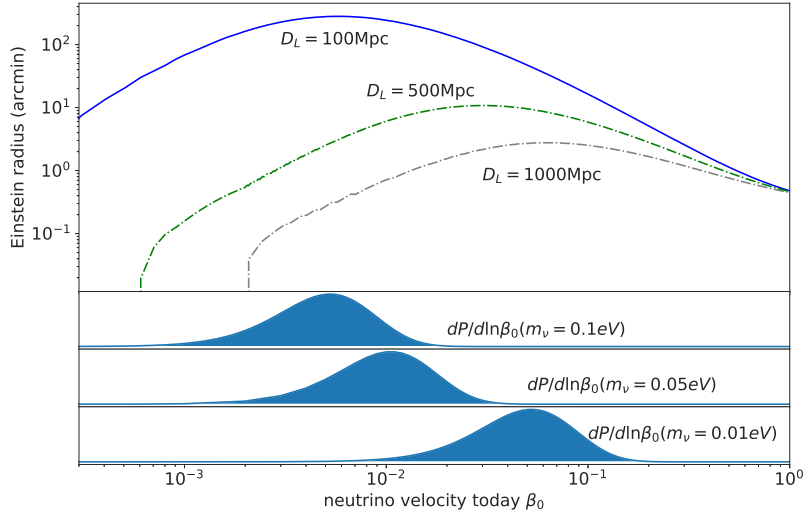}
    \caption{Einstein radii of massive neutrinos passing through  singular isothermal spheres (SIS) with $\sigma_v = 1000$ km/s with $D_L = 100$, $500$, and $1000$ Mpc. The velocity probability distribution follows Fermi-Dirac distribution, we label three specific cases with $m_{\nu} = 0.1$, $0.05$, and $0.01$ eV. The smaller the v, it would cause larger angle of deflection. However, it would also decrease the distance between the lens and the surface of last scattering, hence decrease the delecting angle. For each lens position, it would correspond to a certain $v_0$ that maximize the Einstein Ring.}
    \label{fig:Einstein_radius}
\end{figure}

Figure \ref{fig:Einstein_radius} shows the Einstein radii of SIS lenses at three selected distances $D_L$ . There are several important lensing properties for cosmic neutrinos. 

Neutrinos within a given mass eigenstate have a momentum distribution following the Fermi-Dirac distribution (equation \ref{eq:FermiDirac}). Different momenta, even for the same mass, therefore correspond to different Einstein radii. Hence the lensing of neutrinos could provide rich information about the lens potential.

Massive neutrinos can have a wide range in their distance to their surfaces of last scattering, depending on their velocity today, and possibly very different from that of the CMB \cite{2009PhRvL.103q1301D}. It is possible that a neutrino's surface of last scattering could be closer than the physical position of the gravitational lens (e.g. a galaxy cluster), 
leading to a cutoff in lensing when $D_S = D_L$.

\begin{figure*}
    \centering
    \begin{subfigure}[b]{0.475\textwidth}
        \centering
        \includegraphics[width=\textwidth]{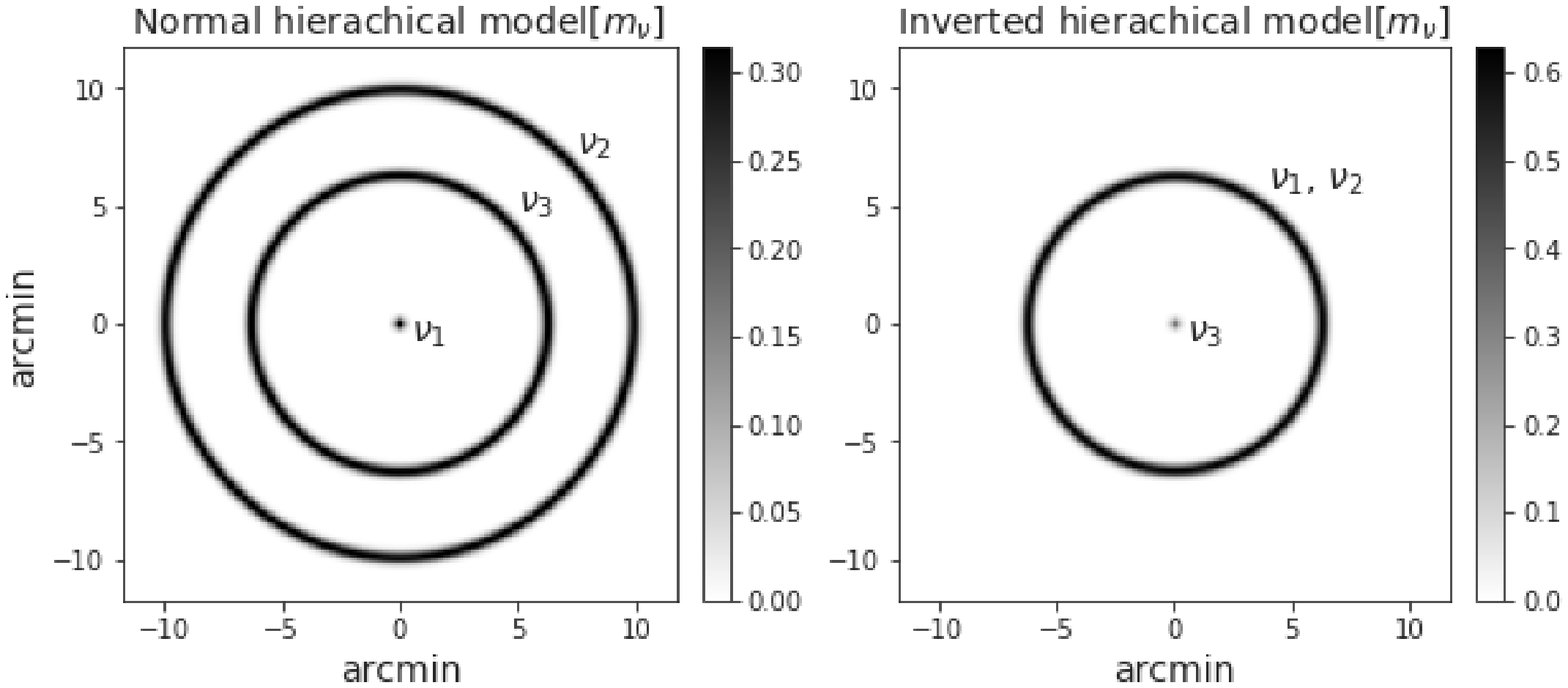}
        \caption[Network2]%
        {{\small Neutrino flux for all mass eigenstates}}    
        \label{fig:mean and std of net14}
    \end{subfigure}
    \hfill
    \begin{subfigure}[b]{0.475\textwidth}  
        \centering 
        \includegraphics[width=\textwidth]{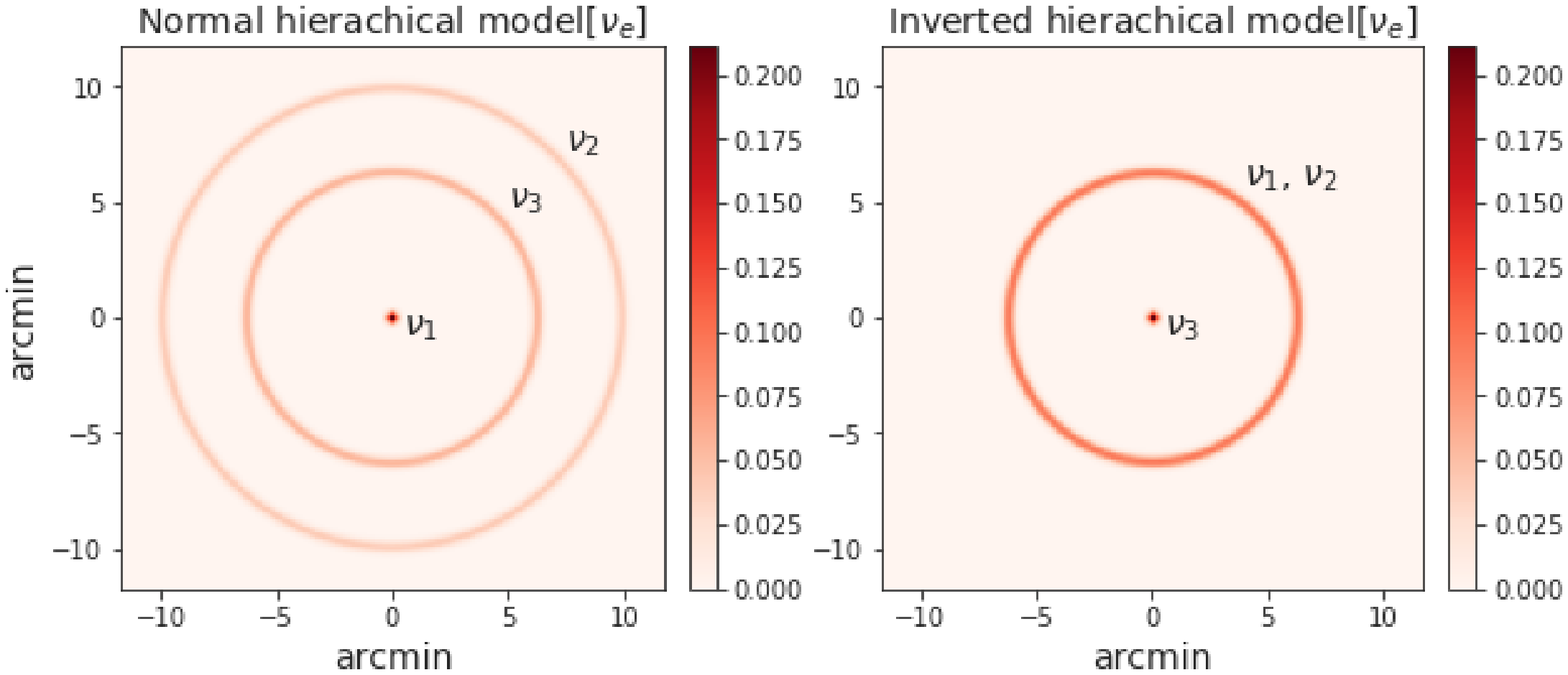}
        \caption[]%
        {{\small Neutrino flux for electron neutrinos}}    
        \label{fig:mean and std of net24}
    \end{subfigure}
    \vskip\baselineskip
    \begin{subfigure}[b]{0.475\textwidth}   
        \centering 
        \includegraphics[width=\textwidth]{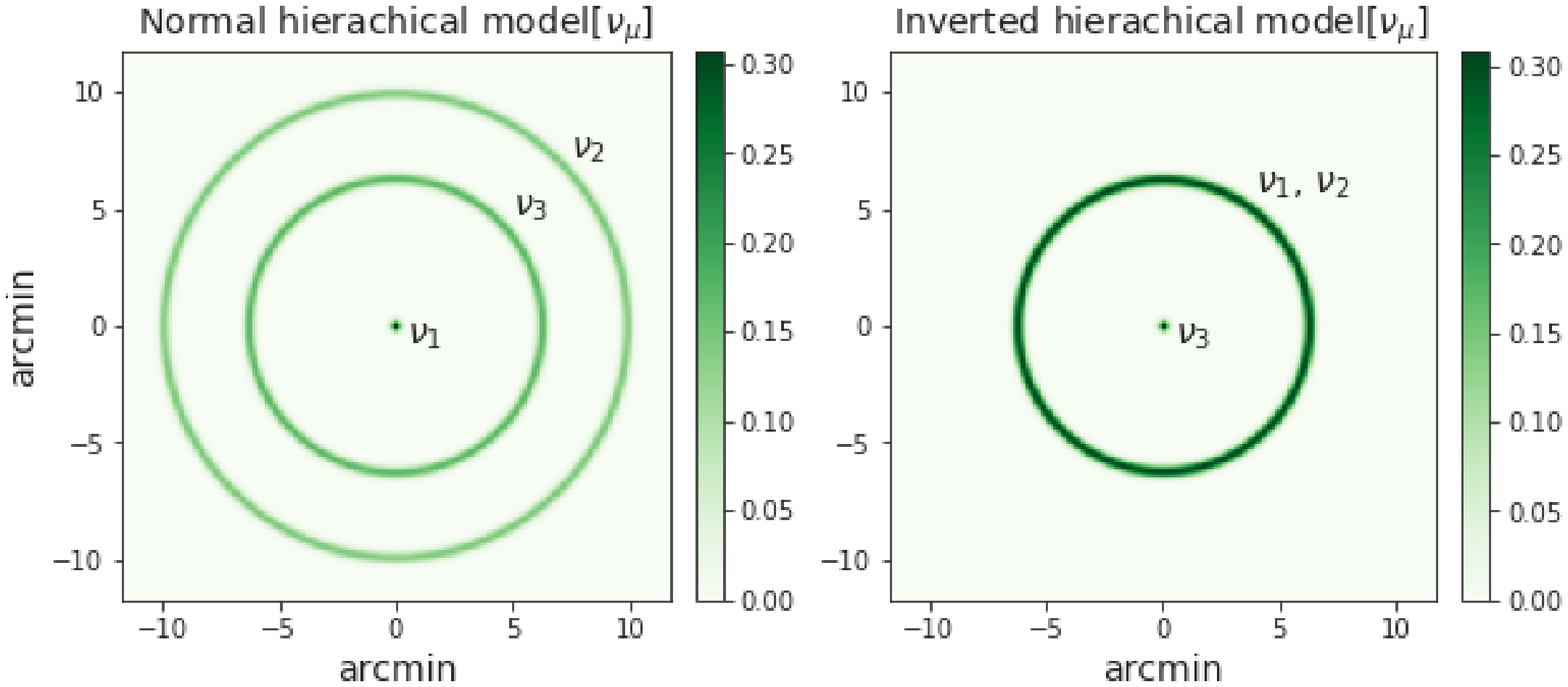}
        \caption[]%
        {{\small Neutrino flux for muon neutrinos}}    
        \label{fig:mean and std of net34}
    \end{subfigure}
    \quad
    \begin{subfigure}[b]{0.475\textwidth}   
        \centering 
        \includegraphics[width=\textwidth]{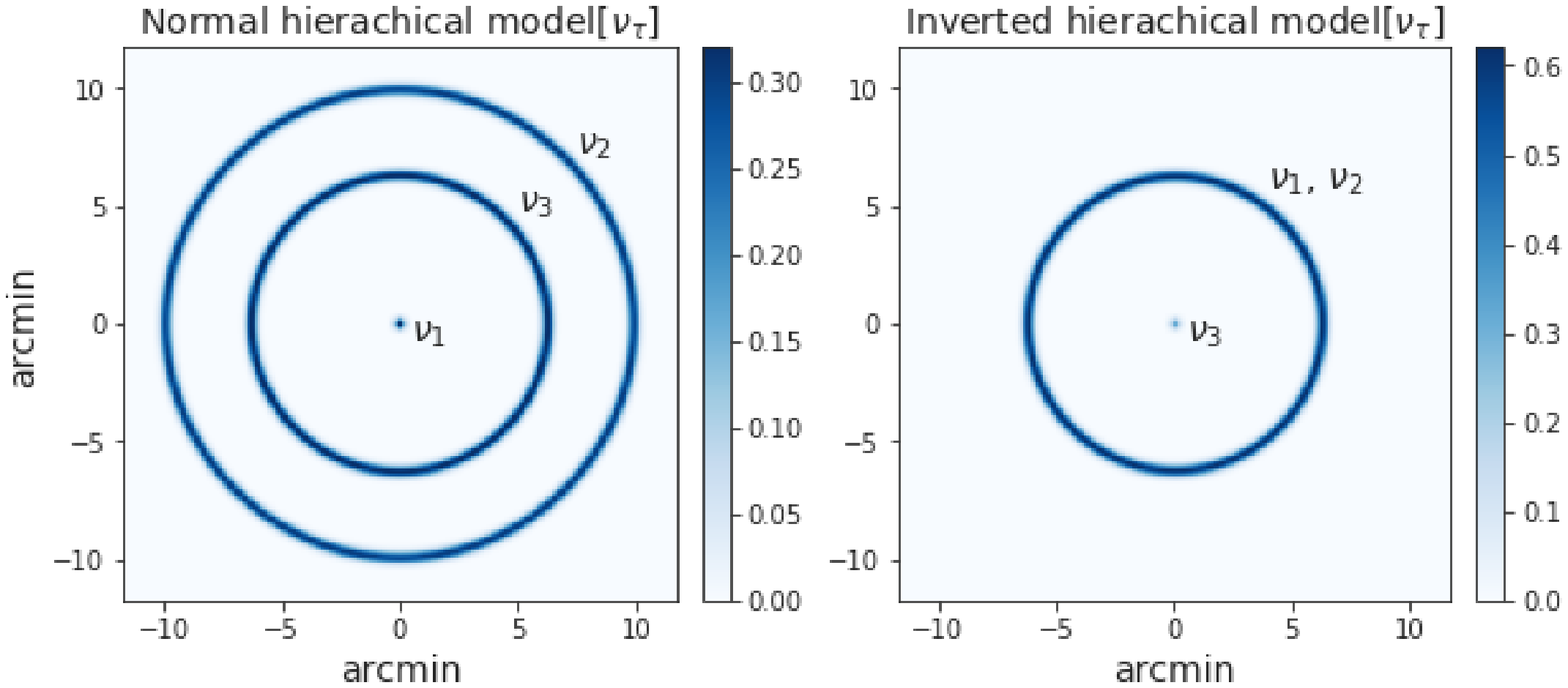}
        \caption[]%
        {{\small Neutrino flux for tau neutrinos}}    
        \label{fig:mean and std of net44}
    \end{subfigure}
    \caption{Lensed image for neutrinos detected as electron neutrinos originating from the line of sight directly through the center of the lens, showing three Einstein rings. The relative fluxes of the rings depend on mixing angles and neutrino hierarchy models: left shows the normal hierarchy, right shows the inverted one, both assuming the lightest neutrino is massless. Here we assume masses of neutrinos $\Delta m_{12}^2 = 8 \times 10^{-5} eV, \Delta m_{23}^2 = 2.32 \times 10^{-3} eV$, and the lightest neutrino mass eigenstate is $10^{-5}$ eV. We are plotting a particular momentum bin, which is the peak of the momentum distribution described by the Fermi-Dirac distribution with the lens is placed at $500$ Mpc away from the observer.}
    \label{fig:3rings}
\end{figure*}

 Neutrinos propagate in mass eigenstates but they are observed in flavor eigenstates, corresponding to a combination of the 3 different mass eigenstates described by the PMNS matrix \cite{doi:10.1143/PTP.28.870,Pontecorvo:1957qd}.  When we observe the lensing pattern for cosmic neutrinos of fixed direction and momentum in a particular flavor state, we are measuring a superposition of three mass eigenstates.  A fixed momentum will correspond to three possible velocities, leading to a superposition of three different lensed neutrino maps, each corresponding to a  different unlensed source plane and a different lookback time to the lens. If the masses and mixings are known, it would be possible to take combinations of lensing maps constructed at different momenta to get the time evolution of individual lenses. For unlensed lines of sight going through the exact center of the halo, there would be three Einstein rings, corresponding to the three mass eigenstates as shown in figure \ref{fig:3rings}. Integrating over momenta, the thickness of individual rings is related to the neutrino momentum distribution following Fermi-Dirac statistics.

Only the lensing of the neutrino background fluctuations can be  detected, as Liouville's theorem ensures that the mean flux is unchanged. The correlation structure of the intrinsically statistically isotropic fluctuations (i.e., there is no preferred orientation on the sky for the fluctuations) is distorted by gravitational lensing. For example, the correlations along the Einstein ring would be strongly oriented along the tangential direction, coming from the lensing of intrinsic fluctuations along the unlensed line of sight that have equal correlations in all directions. Therefore, the lensing effect would look similar to Figure 1 in \cite{hu2002mass}. The Einstein ring is a preferred place in the lensed image where the source-plane isotropic correlations become ring-like in the image. 

We ignore momentum shifts due to decaying or growing potentials (ISW effects) and time delay effects that come from increased time of travel across gravitational potentials leading to the surface of last scattering being at a shorter distance. These are important effects for low-speed neutrinos that should be investigated in future work.

For visualization, we show the lensing of the neutrino background for neutrinos only directly along the line of sight through the center of the lens, as these would have the most dramatic distortion of their correlations, lensed to form Einstein Rings.


\begin{figure}[thpb]
    \centering
    \includegraphics[scale=0.45]{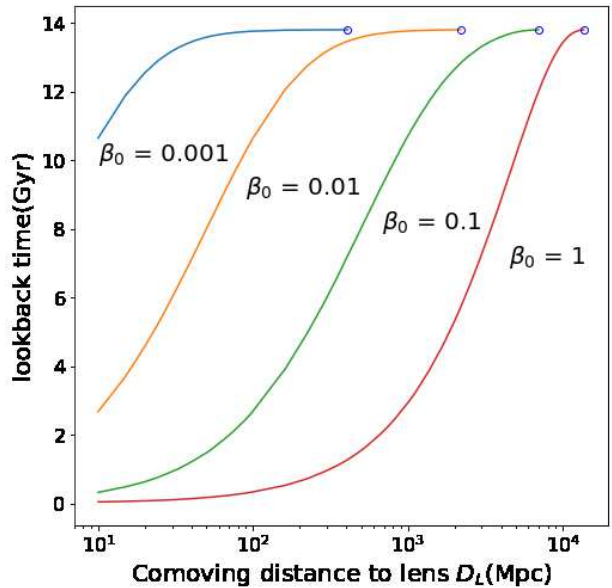}
    \caption{Time (relative to today) that massive cosmic neutrinos pass through a gravitational lens at distance $D_L$, for several different measured velocities today (in units of the speed of light). There is a cutoff when the lookback time approach the age of the universe, as $D_L$ approaches the surface of last scattering $D_S$.}
    \label{fig:lookbacktime}
\end{figure}

The lookback time is the age of the universe relative to today at which a cosmic event occurred, and can be calculated from the scale factor of the universe when the neutrino passed a location a distance $D_L$.
Since neutrinos can travel with different velocities, the lookback time to a lensing event at fixed distance $D_L$ for neutrinos with different velocities can be different. Neutrinos pass the lens at different epochs for the lens. figure \ref{fig:lookbacktime} shows the neutrino lookback time for different velocities today.

If the gravitational lens is evolving, neutrinos with different velocities could be used to measure the time evolution. 
Dark matter halos grow as a function of cosmic time through accretion and repeated mergers. Neutrinos with different velocities today passed through the lens halo at different times. Therefore, they  experienced different deflections due to the mass evolution of the lens halo. In a SIS model, $\theta_E^{\rm SIS} (v_0) \propto \sigma_v^2 (a_{lens})$, where $a_{lens}$ is a function of $D_L$ and $v_0$ described in eqn. \ref{a_lens_D_L_relation}. Neutrinos with larger momentum enter the lens halo later; if the lens halo mass is growing with time, these neutrinos would experience deeper potentials and hence have larger deflections.

\section{Discussion and Conclusion}

Gravitational lensing leads to a breaking of statistical isotropy; the correlations in the neutrino background will have spatial variations induced by lensing, exactly as happens for the lensing of the CMB.  The Einstein ring is the most eye-catching breaking of statistical isotropy by lensing: the unlensed positive or negative fluctuation at a particular momentum will get lensed into a complete annulus of positive or negative intensity relative to the cosmic mean. There will be three such rings for each lens, corresponding to the three mass eigenstates.  Elsewhere in the image, the unlensed fluctuations will similarly have their correlations distorted, but less dramatically.  For simplicity, we focus on Einstein rings to understand the information content of the lensing of the neutrino background. In general, something akin to the quadratic estimator used for CMB lensing \cite{hu2002mass} could be used to extract information from the weak lensing of the neutrino background. The Einstein radius is determined by the mass enclosed in the region at the time when neutrinos at a given speed pass through the lens potential. In this work, we assume the lens potential to be constant for simplicity and focus on the splitting of mass eigenstates for the Einstein rings. Understanding the time evolution of the lens potential and its connection with neutrino lensing will be an interesting extension in future work.

Using neutrinos of a single mass, one could imagine the following program to map out a structure at some fixed distance: mapping the gravitational lensing effects at a fixed momentum (which corresponds for a single mass eigenstate to a particular velocity today) to map the spatial variation of the lens potential, while repeating this at various measured momenta to obtain the time evolution of the lens. The three mass eigenstates lead to three superposed independent versions of this program. While lensing studies with photons give the state of the lens only at a single moment in time, neutrino lensing can reconstruct the full time evolution of gravitational potentials inside the light cone.

Different momenta indeed came from a different radial distance and hence we would probe a different region of the source. Hence, we present the result in individual momentum bins. Averaging over momenta will mix together both the sources (from different line of sight locations) and the lenses (at different times) in a complicated way.

Neutrino oscillations in the usual sense are not directly relevant in this case, as the mass eigenstates, equally populated in the early universe, get dispersed along the line of sight by their different propagation speeds. For neutrinos emitted from a single location in space, the wave packets for each mass eigenstate will be widely separated today. In addition, the lens will also act as a spectrometer, splitting the mass eigenstates in the angular direction. 
Therefore, the probability to measure the lensing-induced correlations in particular flavor eigenstates becomes deterministic.  For example, at a given momentum and flavor there would be three Einstein Rings due to the three mass eigenstates, with the relative amplitudes of each ring set by mixing angles, as shown in figure \ref{fig:3rings} . 

Motion of the lens halos would also affect how neutrinos are deflected. In this paper, we assume all the lenses are in a fixed co-moving frame. However, in reality the foreground halos can have peculiar velocities and therefore induce a slingshot effect. At early time, the neutrinos will be relativistic, so the effect would be small. At late times, neutrinos will have larger deflections, possibly leading to larger changes in the momentum of neutrinos.

These lensing effects can only be seen using the anisotropies in the neutrino background \cite{hu2002cosmic, hu2002mass, lopez1999precision, michney2007anisotropy}. Substantially complicating the interpretation will be the strong late-time evolution of the neutrino fluctuations arising from the linear and non-linear integrated Sachs-Wolfe effects \cite{cooray2002nonlinear}. This is an interesting but challenging project that we leave for future investigations.
 
In summary, we have shown that gravitational lensing of the cosmic neutrino background is in principle an extremely rich source of information, containing the imprint of the gravitational evolution of the entire three-dimensional universe within our causal horizon.

\addtolength{\textheight}{-12cm}   




\section*{ACKNOWLEDGMENT}

The authors would like to thank Sunny Tang, Arka Banerjee, Po-Wen Chang, Mark Chen, and Patrick Draper for useful discussions. This work is supported by Brand and Monica Fortner, and NSF AST-17-15717. GH is a fellow in the Gravity and the Extreme Universe program of CIfAR. This research was supported in part by the Perimeter Institute
for Theoretical Physics. Research at Perimeter Institute
is supported by the Government of Canada through the
Department of Innovation, Science, and Economic Development,
and by the Province of Ontario through the
Ministry of Research and Innovation.

\bibliographystyle{JHEP}
\bibliography{bibliography2.bib}

\end{document}